\documentstyle[12pt,aasms4]{article}

\lefthead{Batuski, et al.}
\righthead{Superclusters in Aquarius}

\begin{document}

\title{Discovery of Extreme Examples of Superclustering in Aquarius}

\author{D. Batuski\altaffilmark{1}, C. Miller, and K. Slinglend\altaffilmark{1}}
\affil{Department of Physics and Astronomy, University of Maine,
    Orono, ME 04469-5709}

\author{C. Balkowski\altaffilmark{1}, S. Maurogordato, V. Cayatte\altaffilmark{1}, and P. Felenbok}
\affil{Observatoire de Paris, DAEC, Unit\'{e} associ\'{e}e au CNRS, D0173,
et \`{a} 
l'Universit\'{e} Paris 7, 92195 Meudon Cedex, France}

\and

\author{R. Olowin}
\affil{Department of Physics and Astronomy, St. Mary's College, Moraga, CA
    94575}


\altaffiltext{1}{Visiting Astronomer, European Southern Observatory, La Silla, Chile.} 


\begin{abstract}
The results of spectroscopic observations of 46 $R \ge 1$ clusters of 
galaxies from the Abell (1958) and Abell, Corwin, \& Olowin (1989 - 
hereafter ACO) catalogs
are presented. The observations were conducted at the ESO 3.6m telescope
with the MEFOS multiple-fiber spectrograph. Thirty-nine of the clusters
lie in a $10^{\circ } \times 45^{\circ }$ strip of sky that contains two 
supercluster candidates (in Aquarius and Eridanus). These candidates were identified 
by a percolation analysis of the Abell and ACO catalogs, using
estimated redshifts for clusters that had not yet been measured.
With our measurements and redshifts from the literature, the
target strip is now 87\% complete
in redshift measurements for $R \ge 1$ ACO clusters with 
$m_{10} \le 18.3$. Seven other clusters
were observed in a supercluster candidate in the Grus-Indus region.
Seven hundred and thirty-seven galaxy redshifts were obtained in these 46 
cluster fields.

We find that one of the supercluster candidates is a collection 
of 14 $R \ge 1$ ACO/Abell clusters with a
spatial number density that is $20\times $ the average spatial  
density for rich ACO clusters.  This overdensity has a maximum  
extent of $\sim 110 h^{-1}$ Mpc, making it the longest  
supercluster composed only of $R\ge 1$ clusters to be identified to date. 
This filament of clusters runs within $7^{\circ }$ of the line of sight in the 
Aquarius region, and on its high-$z$ end, four $R = 0$  
ACO clusters (three of which are $R = 1$ in the Abell catalog) appear to bridge gaps to other clusters, extending the structure to $\sim 150 h^{-1}$ Mpc. 
Our analysis also reveals that another supercluster, consisting of nine rich clusters with an extent of
$\sim 75 h^{-1}$ Mpc, runs roughly perpendicular to Aquarius near its 
low-redshift end.
Both of these superclusters are quite filamentary. Fitting ellipsoids to each
set of clusters, we find axis ratios (long-to-midlength axis) of 4.3 for 
Aquarius and 3.0 for Aquarius-Cetus. Two such filaments in a relatively limited
volume might be surprising, unless rich clusters generally can be found in 
elongated aggregations. We fit ellipsoids to  
all $N \ge 5$ clumps of clusters (at $b = 25 h^{-1}$ Mpc) in the measured-$z$ Abell/ACO $R \ge 1$ clusters sample. The frequency of filaments with axis
ratios $\ge 3.0$
($\sim 20$\%) was nearly identical with that found among `superclusters' in 
Monte Carlo simulations of random and random-clumped cluster samples, however,
so the rich Abell/ACO clusters have no particular tendency toward filamentation. 

The Aquarius filament also contains a `knot' of six $R \ge 1$ clusters
at $z \sim 0.11$, with five of the clusters close enough together to 
represent an apparent overdensity of 150 $\bar{n}$.  There are 
three other $R \ge 1$ cluster density enhancements similar to this knot 
at lower redshifts: Corona Borealis, the Shapley Concentration, and another 
grouping of seven clusters in Microscopium. 
All four of these dense superclusters appear near the point of breaking
away from the Hubble Flow, and some may now be in collapse, but there is 
little indication of any being virialized. With four such objects, studies
of them as a class may now lead to much greater insight into large-scale
processes.

\end{abstract}




%

\section{Introduction}

In the last two decades, several important
discoveries of large-scale structure have been made through 
magnitude-limited redshift surveys of individual galaxies. The first was the
Coma/A1367 supercluster and void combination (Gregory \& Thompson 1978, Tifft
\& Gregory 1988), 
which turned out to be part of the `Great Wall' and `bubble' structures 
of the de Lapparent et al. (1988 and 1991)
CfA redshift survey `slices.' Later came Perseus-Pisces
(Gregory, Thompson \& Tifft 1981), the Hercules supercluster (Tarenghi 
et al. 1979, Chincarini, Rood and Thompson 1981, Gregory and Thompson 
1984), the Local Supercluster (Yahil
et al. 1980, Tully 1982, and the early Center for Astrophysics 
survey work -- Huchra et al. 1983), the Bootes void (Kirshner, et al. 1981),
and the Hydra-Centaurus
supercluster (Chincarini \& Rood 1979), most recently thoroughly mapped
along with the slightly more distant region of the `Great Attractor'
by Dressler (1988 and 1991).
All of these efforts to map using magnitude-limited surveys
of individual galaxies
(and a few other, not complete samplings of some apparent
supercluster regions)
have identified structures of order 30 -- 100 $h^{-1}$ Mpc or more,
but they are limited to volumes only about 100 $h^{-1}$ Mpc deep. (Throughout 
this paper $h = H_0/100$ km s$^{-1}$ Mpc$^{-1}$.)
The Sloan Digital Sky Survey and the 2dF Survey being undertaken in Australia
will, with a few years work, provide redshifts for the $\sim 10^6$ galaxies
needed to map scales of a few hundred megaparsecs with such tracers.
To study such scales now, one needs to sample more sparsely, perhaps 
by using pencil-beam surveys to probe deeply with individual galaxies 
({\it e. g.}, Postman, Huchra and Geller 1986, Kirshner et al. 1987, Broadhurst 
{\em et al.} 1990, Small et al. 1998)
or by randomly sampling galaxies in strips across larger areas of the sky 
as in the Las Campanas survey (Shectman et al. 1996)

Rich galaxy clusters are potentially excellent tracers of mass on the 
larger scales.  With an average spatial separation of $\sim 50 h^{-1}$ Mpc
they are efficient for mapping scales approaching those measured by the COBE 
satellite.  Recent analyses of clusters from the Abell (1958) and Abell, Corwin,
Olowin (1989, hereafter ACO)  
catalogs reveal evidence of three-dimensional superclustering on scales
of 50-100 $h^{-1}$ Mpc ({\it e.g.}, Miller et al. 1998, Batuski \&
Burns 1985b, Batuski et al. 1991, 
Postman, Geller \& Huchra 1986, Postman, Huchra, \& Geller 1992, Bahcall \& 
Soniera 1983, Tully 1987, and Tully et al. 1992). We also note that 
rich (Abell/ACO) clusters are primary constituents of all the superclusters
mentioned in the preceding paragraph. 

However, less than one third of the Abell/ACO 
clusters have measured redshifts (and most of these have only 
one or two galaxies measured (Struble \& Rood 1991b, Postman 
et al. 1992), although recent multi-fiber spectrographic cluster
surveys such as Katgert et al. (1996) and Slinglend et al. (1998) have
started to improve this situation).  
Single-galaxy measurements lead
to substantial uncertainties in the cluster redshifts because of the possibility 
of measuring a foreground or
background galaxy not associated with the given cluster (such projection
effects being important in about 14\% of Abell cluster fields with only
a single previously published redshift measurement, according to Miller   
 et al. 1998), or the
possibility of measuring an `outlier' in a cluster's velocity distribution,
which can 
introduce an error of 500 km s$^{-1}$ or more about 36\% of the time 
(Miller et al. 1998).
The projection problem is especially dangerous for part of 
the region targeted for this redshift survey, because of the very high surface
number density of rich clusters.
Our effort here has been an attempt to minimize these 
projection effects
by collecting 20 or more redshifts per cluster field and then to use
the resulting high-confidence
redshift data to investigate superclusters of Abell/ACO clusters.

We have used MEFOS (Meudon-ESO Fiber Optic Spectrograph, see Felenbok et al.
(1997) for a description)
on the 3.6-m telescope at ESO
to observe clusters in a $10^{\circ } \times 45^{\circ }$ strip of sky that
includes two particularly interesting supercluster candidates (SCCs),
one in Aquarius and one in Eridanus, to determine the significance
of the apparent superclustering. Another nearby (Grus-Indus) 
supercluster candidate was also observed.
Our sample was limited to ACO richness class $R\ge 1$ clusters, with one or 
zero measured redshifts in the literature, 
at high galactic latitude ($\vert b\vert > 30^{\circ }$). 

Our first run of this program took place in August 1994 and the second 
was in September 1995. Two clusters were also observed in May 1995. 
We observed 39
clusters in the Aquarius-Eridanus strip (targeting to get 20 or more redshifts per
cluster field) and seven clusters in the Grus-Indus SCC. We have, to
this point, completed 87\% of the redshift survey of the strip to
$m_{10} \le 18.3$ (estimated $z < 0.16$). We had 47 target clusters in the strip
to this magnitude limit, eight of which had redshifts in the literature, 
and we obtained 
redshifts of 33 more.  Six remain with unmeasured redshifts, including three
that we observed but for which we obtained insuffient data for a cluster velocity
determination. 
A preliminary report on this survey was given in
Slinglend et al. (1995), and the final results are presented here.

Section 2 defines the sample of clusters observed in our primary target strip, 
and Section 3 summarizes
the procedures for conducting the observations and reducing the data. The
cluster redshift results are presented in Section 4 for both the Aquarius-Eridanus 
strip and Grus-Indus, along with  
a discussion of our analyses of the Aquarius superclusters. Section 5 contains 
our conclusions.

\section{Cluster Sample in the Aquarius-Eridanus Strip}

Our entire sample is made up of $R\ge 1$ ACO clusters within a 
$10^{\circ } \times 45^{\circ }$ strip of sky in the southern hemisphere (see 
Fig. 1).  Each point in Fig. 1 represents an $R\ge 1$ 
cluster with $m_{10} \le 18.3$. We used the ACO richness classification for 
consistency throughout the strip, and we note that clusters on the northern 
end of the strip (north of $\delta = -27^{\circ }$) were also in Abell's (1958) 
original catalog, classified in richness with a variable background subtraction
in the member galaxy count, rather than the global background subtraction of 
ACO. The ACO $m_{10}$ magnitude estimations were also used throughout our study.

This particular strip was chosen so as to include 
two apparent overdensities in 
the spatial distribution of rich clusters.  These overdensities (SCCs) were 
identified using a percolation algorithm based upon the 
redshifts from the literature or redshifts estimated from the magnitude of 
the tenth brightest cluster member, using the $m_{10}$ - $z$ relation of ACO. 
Distances were calculated using (Sandage 1975)

\begin{equation}
D = \frac{cz}{H_o}{\frac{(1+\frac{z}{2})}{(1+z)}} 
\end{equation}

\noindent for a Friedman universe with $q_0 = 0$. 
The percolation parameter used to identify these apparent superclusters was
30 $h^{-1}$ Mpc, which corresponds to a spatial cluster density of about
five times the average ACO cluster spatial number density of $8 \times 10^{-6}
h^3$ Mpc$^{-3}$ (Miller et al. 1998).
Since most of the clusters 
had only estimated redshifts, these two tentative supercluster identifications
were considered simply as promising targets in the selection of the strip 
for observing. The clusters in this strip are at high
galactic latitude ($b \sim -50^{\circ }$ to $ -70^{\circ }$), so that 
obscuration is not a concern.  

There has been some discussion in the literature concerning the adequacy 
of the Abell and ACO catalogs of clusters for tracing large-scale structure.
Sutherland
(1988), Olivier et al. (1990) and Sutherland \& Efstathiou (1991)
suggest the impact of redshift-angular separation anisotropies due to
projection effects in the visually selected Abell catalog is severe. 
Struble \& Rood (1991a) show that such
effects are small ($\sim $3\%) among the 1682 clusters in the
`statistical sample' subset of Abell's catalog.  Also, Postman, Huchra
\& Geller (1992) give a strong argument that the spurious structure
due to projection effects is insignificant in the sample of 351 Abell
clusters with measured redshift that they present. On the other hand,
Efstathiou et al. (1992) find significant indications of
artificial anisotropies in the Postman et al. sample and show
that the clusters selected from the APM 
(automatic plate measurement system at Cambridge University) galaxy 
catalog have no such
anisotropies. Bahcall \& West (1992) counter with another analysis,
comparing Abell cluster results directly to the APM work, which
indicated little effect from projection for the Abell clusters.
Recently, Miller et al. (1998) also use newly-expanded samples 
of Abell and ACO $R \ge 1$ clusters with measured redshifts 
to show that anisotropies in the catalogs 
are on the same order as in the APM catalogs and not significant for 
purposes of large-scale studies.
The new sample of Miller et al. consists of a large number (291)
of $R \le 1$ Abell/ACO clusters, 96\% complete to $m_{10} = 16.8$.
This sample is comprised of 198 northern Abell clusters,
(188 of which have more than one measured redshift, with most of these
redshifts from
the MX Northern Abell Cluster Survey of Slinglend et al. 1998) and 91 southern
ACO clusters (most of the redshifts from the ESO Nearby Abell Cluster Survey of 
Katgert et al. 1996).

During our first observing run (August 1994 at ESO), we concentrated on the
most dense of the two
supercluster candidates.  The Aquarius candidate is well-outlined by a 
$10^{\circ }$ box on the sky centered on $\alpha = 23^h.3$, $\delta = -22^{\circ }$ 
(upper left of the $10^{\circ } \times 
45^{\circ }$ strip shown in Fig. 1). We first worked to complete observations
of the previously unmeasured clusters in this smaller region, and during the 
second run (September 1995), covered the larger strip down to our magnitude
limit of $m_{10} \le 18.3$. 


\section{Observations and Data Reduction}

The observations took place over nine nights in August 1994, May 1995 and
September 1995. The observations were under conditions of good to adequate
seeing ($0.5^{\prime \prime}-2.0^{\prime \prime}$) and transparency. The 
instrument used was the MEFOS (Meudon-ESO Fiber Optic Spectrograph) 
mounted on the ESO 3.6m telescope. The grating-CCD combination used  
resulted in a dispersion of 170 \AA /mm and a resolution of about 11 \AA. 
The wavelength range was chosen to be 3800 \AA ~ - 6150\AA.

We collected approximately 15--20 redshift-quality spectra within a
given cluster field per hour. Comparison frames of helium and neon were taken
before each exposure for wavelength calibration. ``Fiber flats" (continuum
spectra) were also taken for use in determining the location of the spectra on
the CCD images.

\subsection{Data Reduction}

The data was reduced in the IRAF\footnote{IRAF is distributed by the
National Optical Astronomy Observatories,
which is operated by the Association of Universities for Research in
Astronomy, Inc. (AURA) under cooperative agreement with the National
Science Foundation.} environment, utilizing standard IRAF routines as well 
as modified routines designed for Steward Observatory's MX Spectrometer.
The MX-specific routines were written by John Hill, Bill Oegerle, David Batuski 
and Kurt Slinglend. The
2-dimensional CCD images were bias-subtracted and the individual spectra were
extracted after being located on corresponding `fiber flat' (quartz lamp) exposures. 
A dispersion solution from each object's comparison (helium-neon) 
frame provided the
spectra wavelength calibrations. A sky subtraction was performed after 
averaging all the sky spectra from a given exposure and normalizing by the 
strength of the [OI] 5577\AA ~night sky line. 

A problem has been reported (Felenbok et al. 1997) with some of the MEFOS fibers
having persistent offsets after wavelength calibration, resulting in shifts 
of $\sim$0.5 \AA ~in the measured wavelength of the 5577\AA ~night-sky line.
We examined our data for this problem and found that seven of our object 
fibers did indeed have such large offsets. Therefore, the
wavelength solutions for galaxies observed with those fibers were each
shifted by the negative of the average offset that we measured for the 
fiber in question.

\subsection{Cross-correlation} 

We used the IRAF task ``fxcor" to perform our cross correlations. Each spectrum 
was first cross-correlated against template spectra of eight stars with 
high-precision published velocities (Maurice et al. 1984). Each of these template stars 
was observed with MEFOS, in single-fiber mode. 
The star spectra were cross-correlated against each other, and the resultant
velocities were compared with published values. Adjustments to our calculated
velocities were then made to better match the published values. These
corrections were on the order of 10 -- 15 km s$^{-1}$ and 
were made to reduce errors in the star template velocities due
to night-to-night instrument variations. 

For an alternate determination of each galaxy's velocity, a bootstrap technique 
was used. That is, the galaxy 
spectra were also
cross-correlated against 20 low-redshift galaxy templates also observed 
on MEFOS (chosen for their high signal-to-noise ratio). The velocities of 
these galaxy templates had been previously determined through cross-correlation 
with the eight star templates mentioned above. In cases of high redshift object 
spectra ($z > 0.1$), the galaxy templates often 
provided a much more reliable velocity determination due to the greater number 
of comparable lines between the spectrum of the object and that of the template.

Each cross-correlation returned a heliocentric velocity and a cross-correlation 
strength (Tonry \& Davis (1979) $R$ value).
For a particular target object, the set of calculated velocities (20 when the 
galaxy templates were used, or eight when the star
templates were used) was then weighted by $(R+1)^2$ and averaged. 
Velocity errors were estimated using   
the $\sigma _v = Q/(R+1)$ method of Hill \& Oegerle (1993) 
and Pinkney et al. (1993) $Q_{MEFOS}$ was determined from a selection of 22
galaxies that were each observed with MEFOS on two different occasions. The 
variations in these velocities were directly calculated, resulting in an average
value of $Q_{MEFOS} = 260$ km s$^{-1}$. Each of the measured velocities was then
assigned an error of $\sigma _v = 260/(R+1)$, where $R$ is now an average value 
(over the successful cross-correlations with the eight star templates or the 20
galaxy templates.) To avoid unreasonably small errors for galaxies with $R$ 
values greater than 13, 
we established a minimum error of 20 km s$^{-1}$. This lower limit allows for
wavelength calibration and other errors in our system, and the magnitude of
the minimum error was estimated from the scatter in measurements of the
5577\AA ~night-sky line, mentioned in subsection 3.1 above.

The results from the star template and galaxy template cross-correlations for
each object were then compared. Cases that did not meet our minimum 
requirements ($R > 2.5$, median absolute deviation (MAD) of the velocities 
found for a galaxy $< 250$ km s$^{-1}$, and a minimum of more than half of the  
templates cross-correlating successfully with the spectrum of the target) 
were discarded. For cases where a reliable redshift was determined 
from both the star template and the galaxy template cross-correlations, a 
comparison was made between the two sets of correlations, considering $R$, 
MAD, and the number of matched templates to determine which ($R+1$)-weighted 
velocity was more reliable. In most cases the velocity from the star template 
cross-correlation matched the velocity of the galaxy templates 
cross-correlation to within 1 $\sigma _v$. In the rare instances where 
these two determinations did not agree, a visual inspection of the galaxy 
spectrum was used to determine which, if either, 
of the  ($R+1$)-weighted velocities was correct. All of the spectra for the galaxies 
listed were examined visually to ensure that the velocities obtained were not 
occasionally the results of fluke occurrences in the data. 

\subsection{Emission Line Redshifts} 
 
All of our sky-subtracted galaxy spectra were examined for emission lines.
If any reasonably likely candidate lines were found in a spectrum, the 
redshift of the galaxy was calculated using the rvsao.emsao package provided 
by the Smithsonian Astrophysical Observatory Telescope Data Center as an
add-on to IRAF.  If a galaxy had two or more apparent emission lines, all the  
good (not distorted by cosmic ray strikes, too low signal-to-noise, etc.) 
lines were used in emsao for the calculation. If only one emission line was 
found, and an absorption line velocity had been obtained that was in rough
agreement with the emission line velocity, the single line velocity was
accepted. A single emission line velocity was also considered 
acceptable even when the absorption line velocity for a galaxy was not deemed
good enough for publication, if there were absorption features in good 
agreement with the location of the emission line.

The emsao task calculates a mean velocity by weighting the velocities returned
by each emission line fit where:
\begin{equation}
W_{i}=\frac{\frac{1}{dvel_{i}^{2}}}{\sum 
\frac{1}{dvel_{i}^{2}}} 
\end{equation}
where $dvel_{i}$ is the error in the fit of the emission
line with an RMS dispersion of $.05\AA$.
Emsao returns similarly weighted errors. In addition to the line fit
errors, we have found a 16 km s$^{-1}$ root mean square variation
when comparing
the 5577\AA ~lines locations (see Section 3.1). 
Total errors in the emission line velocities were calculated by
the root sum square of the error in the line fit and this inherent
scatter (in the night-sky line wavelengths) due to the instrument. 
For the few galaxies where only the OII$_{3727}$ was detected, a third
systematic error was added because of the lower accuracy of the  
wavelength calibration towards the blue end of the
spectra. This additional error was determined to be 
22 km s$^{-1}$ by comparing the velocity determination
using only the OII$_{3727}$ line with the velocity using
multiple lines. We used 18 galaxies with multiple emission lines, that included 
OII$_{3727}$, for the calculation of this third error.

\section{Results}

The recession velocities of all the galaxies observed in this program are 
presented in Tables 1, 2, and 3. Table 1 lists the results for galaxies in 
the MEFOS fields of clusters in the $10^{\circ }$ square region of sky
centered on $\alpha = 23^h.3, \delta = 
-22^{\circ }$, the heart of the Aquarius supercluster candidate.
Table 2 gives the galaxy data for the clusters in the remainder of the target 
strip shown in Fig. 1. Table 3 presents the results for another supercluster 
candidate (Grus-Indus) that we had the opportunity to observe.
The first two columns in each of these tables are the right ascension and 
declination coordinates (J2000) of the galaxies observed. Column 3 contains the 
 ($R+1$)-weighted heliocentric velocities, and Column 4 of each table 
lists the estimated errors ($\sigma _v$). The last column
has an object number entered if the subject galaxy has an emission
line velocity (for cross-reference to Table 4).
Velocities and errors given 
in Tables 1, 2, and 3 were calculated from the absorption lines in the spectra,
unless there is an entry in the last column (Emission Reference) that ends
with the letter `e.' In those cases, the emission line velocity is listed 
in the Tables 1, 2 or 3. Entries in the last column that end with the letter
`a' indicate that an emission line velocity was determined (and is listed in
Table 4), but the absorption line velocity is listed in the Tables 1, 2 or 3 and
was used in our subsequent study of structure in the region.


Table 4 lists the emission line velocities for all our target galaxies 
that had acceptable emission lines. The first column indicates the Abell/ACO
cluster, while the second column lists the reference number from Column 5 in
Tables 1, 2 
and 3. The third and fourth columns list the emission line velocities and
estimated errors.
The fifth column indicates which lines were detected in the spectrum. Lines
within parentheses were seen, but not used in the velocity determination.

We obtained 737
redshifts in 46 cluster fields in our program. Thirty-nine 
of those cluster fields were in
our $10^{\circ } \times 45^{\circ }$ strip. 
Nineteen additional  $R \ge 1$ clusters in 
the strip had previously been observed, including three by Batuski et al. 
(1995) and eight     
with a single redshift per cluster by Ciardullo et al. (1985).
The velocities of these clusters are presented in Table 5. Tables 5a, 5c and
5d list the Abell/ACO cluster number in column one. The mean and 
standard deviation are in columns two and four. 
Columns three and five labeled $C_{BI}$ and $S_{BI}$,
correspond to the bi-weight estimate of the data's central location and scale
(dispersion) respectively. These estimators were calculated using the ROSTAT
statistical analysis package developed by Beers, Flynn \& Gebhardt (1990).
All velocities reported have been corrected for heliocentric motion.
The final column indicates the number of galaxy redshifts that went into
the velocity determinations. Table 5b lists fourteen clusters that are in the 
$10^{\circ }$ square containing the Aquarius supercluster, but
observed by other researchers.


Insufficient data were collected on clusters A2641, A3842, and A3861 
to make a velocity determination. Clusters A3725 and A3944
were determined not to be clusters at all, but simply many galaxies
strung out along the line of sight. These two clusters have been dropped 
from our sample, although, since A3725 has $m_{10} = 18.5$, it did not affect 
our level of completion within the strip.
 
Mean cluster velocities for the observed Abell clusters in this program have
been determined with a procedure that has grown from a 
compilation of ideas of previous investigators (e. g., Batuski and Burns, 
1985a, Postman, Huchra \& Geller, 1992, Slinglend 1996). The data 
for each cluster field was examined for a grouping of four or more galaxy 
velocities with no gaps greater than 900 km s$^{-1}$, to serve as a starting
point for an iterative membership determination. These
velocities were used to calculate the classical mean and standard deviation,
as well as a bi-weight location and scale of the cluster. These locations
and scales were determined using the ROSTAT package
(Beers, Flynn, \& Gebhardt, 1990).  After the mean and standard deviation 
were calculated, any galaxies of the initial grouping that did not fall 
within 3$\sigma $ of the mean velocity are excluded from cluster membership 
for the next iteration of the classical statistics calculation. Likewise, 
data points just on the other side of a 900 km s$^{-1}$ gap were added for 
the calculation, as long as they did not fall outside 3$\sigma $ from the 
(newly recalculated) mean velocity.    In some
cases, there are simply too few galaxy velocities in the cluster to do a
statistical analysis. When N = 3 (and even N = 2 in the case of A3205, where cluster 
membership, based on relative distance from the projected cluster center, 
appeared very likely), only a simple mean is calculated from 
galaxies that fall within a $\pm $3000 km s$^{-1}$ grouping. In these cases, 
the calculated mean should be interpreted with caution.

In general this program did not observe Abell
clusters for which redshifts had been previously
measured. However, most of the previously-measured clusters 
within the Aquarius region have only a single measured redshift, and four of
these were observed in our program for comparison,
as well as confirmation of the redshifts.
A2547, A2548, A2555, A2556 were studied by 
Ciardullo et al. (1985) at CTIO and observed again 
for this paper.  These four were chosen because they constitute two pairs
with small angular separation and each pair could be observed during a single 
exposure. Note that different galaxies were
measured for this program than the ones measured by Ciardullo et al.
The velocities for A2547, A2548, and A2556 that were determined by
Ciardullo et al. (1985) were each well within the typical dispersion of
Abell clusters (about 750 km s$^{-1}$, see for example, Zabludoff, et al. 
1990) of the mean velocities calculated  
from our observations. On the other hand,
Ciardullo et al. (1985) determined a velocity for A2555 
of 41550  km s$^{-1}$ by measuring one redshift and we obtained
a mean cluster velocity of
33110 km s$^{-1}$, based on three galaxy redshifts.

\subsection{The Aquarius Superclusters} 

One of the particular supercluster candidates (located in Aquarius)
we investigated in the August 1994 and September 1995
ESO runs is one that was originally identified by Ciardullo et al.  (1985).
Ciardullo et al. identified the supercluster candidate via a 
``friends of friends" percolation scheme very similar to the one discussed
in Section 2.  
They then measured single redshifts for 
20 $R \ge 0$ Abell clusters in this region of sky and, from their results, 
concluded that there was no
supercluster, only a chance projection of clusters along the line of
sight out to $z \le 0.2$.  However, as pointed out earlier, measuring only a single
redshift is dangerous, particularly in a region of such high surface density.
We similarly (via percolation, see Section 2) identified a more 
extended version of the same supercluster candidate.  Our
identification included 39 clusters, 20 of which were originally identified by
Ciardullo et al. (1985). The additional clusters from our analysis were, in 
projection, apparently farther outside the densest clump of clusters in Fig. 1
than the targets of Ciardullo et al. (1985).

The major finding to emerge in our analysis of this data is the confirmation of 
the presence of two extremely large and dense superclusters in Aquarius. 
Fig. 2a is a wedge plot of the redshift distribution for all $R \ge 1$ 
clusters that now have at least one measured redshift within our 
$10^{\circ } \times 45^{\circ }$
strip of the sky.  The triangles and circles in this figure represent the clusters with 
$m_{10} \le 18.3$ (on the ACO scale, which, largely because of the different
media used, differs from the Abell 1958 magnitude scale), which are 87\% 
complete, with 41 out of 47 clusters measured. The circles indicate clusters with
a single measured redshift.
The crosses in this figure are a few fainter clusters that had 
redshifts in the literature or
that we observed 
because more time was available each night for the ends of the strip. Asterisks
in this plot represent $R = 0$ ACO clusters with measured redshifts. As mentioned
below, several of these $R = 0$ clusters were classified as $R \ge 1$ in Abell
(1958).
Fig. 3 is the same as Fig. 2, except that $R = 0$ clusters are not
plotted and Abell/ACO catalog numbers have been added.

The region from $cz = 24000$ to $cz = 36000$ km s$^{-1}$ 
(an extent of $\sim 110 h^{-1}$ Mpc) around $\theta = 10^{\circ}$ in Fig. 3
contains 14 Abell/ACO clusters that percolate into a single structure with a 
percolation parameter of length $b = 25 h^{-1}$ Mpc, which corresponds to a 
spatial number density of 
clusters, $n$, that is eight times the average spatial number density of ACO 
clusters, $\bar{n}$ (see Section 2). In a percolation analysis at this density 
threshhold of all 
the Abell/ACO $R\ge 1$ clusters with measured redshifts (similar to that of
Bachall \& Soniera 1984 and Batuski \& Burns 1985a, but including redshifts 
from Slinglend et al. 1998, Katgert et al. 1996 and other 
sources in the literature), this was the structure of greatest extent
that appeared. It was also one of the two structures with the largest number 
of member clusters. The Shapley Concentration percolates
at this density to also include fourteen $R \ge 1$ clusters with 
an extent of only $56 h^{-1}$ Mpc, while Corona Borealis becomes a complex 
of ten clusters with a $60 h^{-1}$ Mpc extent. Two other supercluster
complexes, each with six member clusters, have longest dimensions of
$60 h^{-1}$ Mpc in this analysis, and one structure (discussed below)
consists of nine clusters in a $75 h^{-1}$ Mpc filament. We note also 
that the Perseus-Pegasus supercluster filament (Batuski \& Burns 1985b)
and several other previously identified superclusters did not show up in this 
analysis since they are composed of a substantial number of $R = 0$ clusters.

The value of $n = 8\bar{n}$ is essentially the minimum value within the 
supercluster, since it corresponds to the value of the percolation 
length that connects the structure through its sparsest regions.
The average density
of this supercluster can be estimated by calculating the volume in a 
near-rectilinear box that encloses all the 14 clusters within their extremes of 
$\alpha, \delta$, and $z$. Such a box has a density that is a
factor of 20 times the average $R \ge 1$ ACO cluster
spatial number density.

Of the 14 clusters in the filament, four (listed in Table 5b 
and identified with open circles in Figures 2a and 3)
still have only a single measured
redshift, and as we mentioned in Section 1, this increases the 
risk of problems due to projection effects. However, only two of these 
single-redshift clusters, A2528 and A2541, are in `bridging' 
locations within the Aquarius filament, such that new, significantly
revised redshift information would cause substantial break-up of the
percolated filament at $b = 25 h^{-1}$ Mpc.
With about 14\% of Abell clusters with single reported 
redshifts turning out to have substantially revised redshifts after
multiple measurements (see Section 1), there is thus a roughly 25-30\%
chance that additional observations will have large
negative impact on the filamentary structure we report here. On the other hand, 
the region would still contain an unusual amount of superclustering (and if a cluster's
redshift is revised, it is still quite likely to be part of that structure given the 
line-of-sight nature of things in this region). Additionally, there are
six single-measurement clusters in Table 5b that do not currently 
appear to be part of Aquarius, and about 20 fainter
clusters with no measurements that could be found to occupy or extend
this filament upon further observation.


There is an especially tight knot of six clusters
in the region near $cz = 33000$ km s$^{-1}$ and 
$\theta = 10^{\circ}$ in Fig.s 2a and 3 
(clusters A2546, A2553, A2554, A2555, A2579, and A3996). 
These clusters percolate with $b = 15.7 h^{-1}$ Mpc ($n = 32 \bar{n}$),
and without A2553, the remaining five hold together at 
$b = 12.5 h^{-1}$ Mpc ($n = 64 \bar{n}$). If their redshifts are 
directly translated into distance according to Equation (1),
this group of five occupies a sphere with radius $10 h^{-1}$ Mpc (allowing 
sufficient room for the clusters to `fit' by adding 2 $h^{-1}$ Mpc 
to the minimum radius that would barely contain the apparent cluster 
center positions). The sphere would then  
have a spatial number density about 150 times the average for $R \ge 1$ ACO 
clusters, an amazing result, comparable to the five-cluster, densest 
portion of Corona Borealis 
supercluster (Postman, Huchra \& Geller 1986), which is about 100$\bar{n}$, 
and the nine $R \ge 1$ clusters of the Shapley Concentration 
(Scaramella et al. 1989) that percolate at $b = 10 h^{-1}$ Mpc
to have $\sim 110\bar{n}$ for a similar spherical volume. 

Only one of the clusters in the Aquarius knot, A2546, is limited to 
a single measured galaxy redshift, so high confidence can be placed 
on the reality of this extreme density peak (as is also the case for 
the three other peaks discused above).
Note also that A2548, an $R = 0$  cluster that was included on our list
of targets because it could be observed in the large MEFOS field at the 
same time we were observing A2547, also happens to have a redshift 
that places it in the knot, contributing further to the density of the region.
A2548 had been classified $R = 1$ by Abell (1958).

On the high redshift end of the Aquarius filament, four $R = 0$ (by ACO
criteria) clusters are in positions that bridge gaps at $b = 30 h^{-1}$ Mpc
(see the asterisk symbols in Fig. 2a), further extending the structure 
to a total length of 150 $h^{-1}$ Mpc, by connecting in the $R \ge 
1$ clusters A2521, A2547, A2550, and A2565.
Three of these clusters, A2518, A2540, and A2542, had been classified as
$R = 1$ in Abell's catalog and were reclassified as $R = 0$ in ACO. The other 
cluster, A2568, is classified $R = 0$ in both catalogs. Redshifts for all 
four of these clusters were measured by Ciardullo et al. However,
since the $R = 0$ cluster redshifts are very incomplete in this region (for
instance, only five out of 16 measured in the $10^{\circ } \times 10^{\circ }$ 
square of sky around the projected Aquarius filament), no analysis of the significance
of their contribution to the structure in the region is appropriate at this
time.

There is another collection of nine ACO/Abell $R \ge 1$ clusters that percolates 
at $b = 25 h^{-1}$ Mpc ($n \sim 8\bar{n}$) across the foreground in the region of 
the Aquarius supercluster. The member clusters are A2456, A2480, A2492, A2559, A2569, 
A2638, A2670, A3897 (also listed as A2462 in Abell 1958), and  A3951. Since 
A2670 is near the Aquarius-Cetus boundary, we will hereafter identify this 
supercluster as Aquarius-Cetus, to distinguish it from the Aquarius supercluster 
discussed above. The wedge plot in Fig. 4 is centered on the dashed line in Fig. 1, 
and with its $12^{\circ }$ width, it covers both of the superclusters in this
region. The clusters of the Aquarius supercluster are identified with filled
circular symbols, while Aquarius-Cetus clusters are filled triangles. Open triangles
stand for clusters that do not percolate into either supercluster at $b = 25 h^{-1}$ Mpc.
The Aquarius-Cetus complex spans an overall distance of 75 $h^{-1}$ Mpc, and 
it also has an especially high-density
group of four clusters (A2456, A2480, A3897, and A3951), with an average
density (sphere-average method above) of 70$\bar{n}$. All four of these clusters
have multiple measured redshifts.


At the relatively low density threshhold of $n \sim 4\bar{n}$, 
the Aquarius and Aquarius-Cetus superclusters percolate together, through 
the crossing of the gap between A2538 and A3951.

Another approach to analyzing our redshift data for the Aquarius region 
is to consider our 
observations as $1^{\circ }$ wide pencil-beams sampling this small region of
the sky in several spots (especially since one is compelled,
with the large field and the mechanical arms of MEFOS, to chose several target 
galaxies per field that are well away from the apparent cluster center). We can 
then view the aggregate velocity distribution of the galaxies observed in 
this region with histograms (Figures 5 and 6).
As one might expect from the above discussion of the knot of six clusters, 
Fig. 5 shows a large peak at $\sim 33000$ km s$^{-1}$. However, when the 
velocities associated with the clusters in each of our fields are subtracted, 
we have the result in Fig. 6, which shows a
sizeable (about 20 galaxies) peak in the velocity distribution remains at 
$\sim 33000$ km s$^{-1}$. This suggests that several of the fields targeted 
toward other clusters might be sampling a more extended `background' population 
around the six cluster knot (as well as another population around 26000 km s$^{-1}$). The 2dF survey of this region should confirm 
whether or not such extended structure exists around the knot of clusters.



We also point out 
that the percolation analysis of $R \ge 1$ Abell/ACO clusters with measured 
redshifts adds a seventh cluster (A3677) to another high-density 
(percolating at $b = 12.5 h^{-1}$ Mpc) supercluster in Microscopium identified
by Zucca et al. (1993) as consisting of six measured-redshift $R\ge 1$ clusters 
(A3682, A3691, A3693, A3695, A3696, A3705). Each of these seven
clusters has many measured redshifts from Katgert et al. (1996).
Fitting of a sphere around the five of these clusters that percolate
at $b = 10 h^{-1}$ Mpc results in $n = 130\bar{n}$. Of course, the densities 
calculated by this sphere-fitting algorithm may differ considerably from
the actual spatial densities of the clumps considered, if there is much
of a peculiar velocity/dynamical component to the redshifts of the clusters
(see section 4.3 below),
but treating these apparent peaks in the same way allows for a rough
estimation of their comparative densities. Thus, including Shapley and Cor Bor, we have
four very high density peaks, each involving at least five rich clusters, 
within $z \le 0.11$. 

\subsection{Significance of the Superclustering in Aquarius} 

We took two approaches to evaluating the statistical significance of the 
Aquarius supercluster filament. Both of these involved analysis of simulated
`universes'
of point locations in space corresponding to the centers of pseudo-clusters 
of galaxies with two-point spatial correlation
functions very similar to that of the $R \ge 1$ Abell clusters. Following 
the procedure of Batuski \& Burns (1985b), which was based on the hierarchical
nested-pairs technique of Soneira \& Peebles (1978), one hundred catalogs of 
pseudo-clusters were created by placing pairs of pairs within
a spherical volume in such a manner that $\xi (r)$ closely matched that of
the recently enlarged sample of Abell clusters with redshifts analyzed in Miller
et al. (1998), as shown in Fig. 7. This procedure can be visualized
as placing a `long rod' at a random point in space with a random orientation,
then at each end of the rod placing a short rod, again with random orientation,
and finally placing a cluster at each end of each short rod,
dropping any clusters that were outside the radius of 
the spherical volume of the simulation. The lengths of the rods were governed by 

\begin{equation}
\Lambda = A(1 - Bx^{\alpha}), 
\end{equation}

\noindent where $A = 22 h^{-1}$ Mpc and $B = 0.5$ for the long rods, $A = 11 
h^{-1}$ Mpc and $B = 0.96$ for the short rods, $\alpha = 1.5$ for both
cases, and $x$ is a uniformly distributed random number between 0 and 1. 
Finally, 25\% of the points generated with this algorithm were randomly selected
for deletion from the models so that $\xi (r)$ on small scales 
(5 -- 10 $h^{-1}$ Mpc) closely follows the Abell/ACO case and so that the 
ratio of triple-cluster clumpings to double-cluster clumpings (about 7:1) 
could begin to approach that observed in the Abell/ACO
catalogs (about 3:1) at $b = 25 h^{-1}$ Mpc. These simulations thus have
large-scale clumping very similar to that of the Abell/ACO clusters but no 
particular tendency to form filamentary structures, since the spatial
orientations of all the pairs were determined randomly (uniform in $\phi $ in 
the $x-y$ plane and uniform in $cos \theta $ for the angle with the $z$-axis). 


\subsubsection{Two-dimensional Density Peaks} 

We first examined each of the 100 pseudo-cluster catalogs from the central
position within its spherical volume of space to find peaks in the projected
2-D number density of clusters on the `sky' of an imagined observer at the center 
of such a universe. In the $10^{\circ } \times 10^{\circ }$ square of 
sky centered on the densest part of the Aquarius supercluster 
($\alpha = 23^h.3, \delta =  -22^{\circ }$), there are 23 $R \ge 1$ 
Abell/ACO clusters to a depth of 400 $h^{-1}$ Mpc,
only one of which has an unmeasured redshift (estimating the redshift from
the ACO $m_{10} - z$ relation, the 400 $h^{-1}$ Mpc limit corresponds to 
an ACO magnitude limit of $m_{10} < 18.3$ and an Abell magnitude limit of 
$m_{10} < 17.6$). This is the most pronounced peak surface density of measured 
and unmeasured $R \ge 1$ clusters to this depth on our sky.
One might think that part of this effect is the result of this region being 
so complete in redshifts, but even counting more deeply, using 500 Mpc or 600
Mpc as the cutoff,
beyond which distances there are few clusters with measured $z$ so that 
effectively there is only a uniform magnitude limit, this peak remains 
the highest in our sky by at least 15\%.

In the 100 psuedo-cluster simulations, surface density peaks of similar
amplitude (22 - 25 clusters in a 10$^{\circ}$ square) were identified on
the `sky' of our central observer, and then the clusters within that square
were checked to see how often they would percolate, with $b = 25$ $h^{-1}$ Mpc, 
into a structure of length 100 $h^{-1}$ Mpc or greater. This happened in 
about 1\% of the peak-surface-density cases, so having such a long structure 
roughly along the line of sight is a 
two sigma event in a population of pseudo-clusters that are clumped in a 
scheme that closely approximates the two-point correlation function of
rich Abell clusters but avoids the introduction of more than chance filamentation. 
This result suggests that finding such a structure as the 
Aquarius filament through the observational approach that we used is very 
unlikely unless rich clusters in the 
real universe are more commonly members of filamentary supercluster structures
in comparison to the pseudo-clusters in the simulations.

\subsubsection{Ellipsoid Fitting} 

In the second analysis of the significance of the Aquarius supercluster,
we first fit a triaxial ellipsoid through 
the 14 member cluster positions, using the technique of Jaaniste et al.
(1998). The resulting ellipsoid had axes ratios of 4.3:1.0:0.70, with the 
long axis
tilted at $6.9^{\circ }$ from the line of sight. For examining filamentation
in the Abell/ACO catalogs and the 100 pseudo-cluster catalogs, we chose to 
classify as filaments supercluster ellipsoids with a long axis at least 
three times the length of the longer of the other two axes (axis ratio of 
$R_A \ge 3.0$), so that such identified filaments could have some substantial 
curvature
as well as one axis that was clearly much longer that the others.  

The Abell/ACO sample that we examined for filamentation includes
all $R \ge 1$ clusters with galactic latitude at least 30$^{\circ }$ from the 
galactic plane, measured redshifts, and distances less than 300 $h^{-1}$ Mpc, 
using Equation (1). The sample includes 50 new cluster redshifts soon to be
published in Miller et al. (1999).
The distance cutoff was chosen because Miller et al. (1998) show that
the $R \ge 1$ Abell clusters north of $\delta = -17^{\circ }$ and the 
$R \ge 1$ ACO clusters south of that declination each have a relatively flat 
spatial number density distribution out to $z = 0.10$ before it drops off
steeply, indicating that redshift coverage is quite complete within 
$D \sim 300 h^{-1}$ Mpc. In this sample of 370 clusters, using a percolation 
length of $b = 25 h^{-1}$ Mpc, as we used above for defining the Aquarius and 
Aquarius-Cetus superclusters, there are 64 superclusters (with two or more member
clusters), 14 of which having five or more members. One of these superclusters
consists of six of the clusters found to be members of the Aquarius 
supercluster (the other eight Aquarius member clusters are more distant than 
$300 h^{-1}$ Mpc and were thus excluded from the sample), and another is 
Aquarius-Cetus. 

When we fit ellipsoids to the
superclusters with five or more members ($N_c \ge 5$ at $b = 25 h^{-1}$ Mpc), 
only three among these 14 superclusters satisfied our definition of a filament: 
Ursa Major (with $N_c = 5$) had $R_A = 4.7$, 
Virgo-Coma ($N_c = 6$, none of which is the Coma Cluster) 
had $R_A = 3.2$, and Aquarius-Cetus ($N_c = 8$) had $R_A = 3.0$. 
Ursa Major runs closest to the line of sight, with 
a tilt of only 14$^{\circ }$, while Virgo-Coma and Aquarius-Cetus are tilted
68$^{\circ }$ and 73$^{\circ }$, respectively.
The ellipsoid fit to the fragment 
of the Aquarius supercluster that lies within $D = 300 h^{-1}$ Mpc had
$R_A = 2.5$, so it
did not qualify as a filament.

We then searched for filaments in fifty catalogs of pseudo-clusters 
with $\xi (r)$ constrained
to approximate that of Abell/ACO clusters. Limiting these 
catalogs to the same distance and galactic latitude ranges as
the Abell/ACO case above, there were an average of 440 pseudo-clusters 
per sample. This number is somewhat greater than the 370 Abell/ACO clusters
in our sample, primarily because we chose to use the average space number 
density of ACO clusters ($\sim 8 \times 10^{-6} h^3$ Mpc$^{-3}$) throughout
these pseudo-cluster catalogs rather than the density of Abell clusters
($\sim 6 \times 10^{-6} h^3$ Mpc$^{-3}$, Miller et al. 1998) or an average of 
the two densities. We chose the 
ACO density because we have been using ACO criteria throughout the 
characterization of the Aquarius supercluster. We also point out that
the parameter of interest here is the {\it fraction} of $N_c \ge 5$ 
superclusters that have $R_A \ge 3.0$, which is not affected by a small change
in density.

With $b = 25 h^{-1}$ Mpc, each of the pseudo-cluster catalogs percolated
an average of 7.7 superclusters with five or more members, and 20\% of 
these superclusters satisfied the $R_A \ge 3$ definition of a filament.
(For reference, the same analysis of 100 catalogs of entirely random cluster
positions yielded 0.64 $N_c \ge 5$ superclusters per catalog, 19\% of which
were filamentary, indicating that our pseudo-cluster creation algorithm did
indeed not introduce extraneous filamentarity.) 

The surprising conclusion to this analysis is that even though filamentary
arrangements of galaxies and clusters of galaxies (generally including $R=0$
clusters) have commonly been reported in observational studies of large-scale
structure ({\it e. g.}, Batuski \& Burns 1985b, de Lapparent et al. 1988 and 
1991, Giovanelli \& Haynes 1993, da Costa et al. 1994) and even though this 
one region of the sky in Aquarius contains two pronounced filaments of rich clusters,
the large and nearly complete sample of $R \ge 1$ Abell/ACO clusters actually
shows no more filamentation than what could be expected by chance alignments
in a clumpy universe. While the Abell/ACO sample
contains a much higher number of large ($N_c \ge 5$) superclusters than the 
average pseudo-cluster catalog (14 versus 7.7), essentially the same fraction
of such superclusters in either case (21\% for Abell/ACO and 20\% for 
pseudo-clusters) satisfy $R_A \ge 3$. For $b = 20$ and $30 h^{-1}$ 
Mpc, and for $R_A \ge 2$ as well as $R_A \ge 3$ for the filamentation 
threshold, the Abell cluster
sample also showed similar amounts of filamentation to that seen in 
the pseudo-cluster catalogs.

We thus appear to have a contradiction between our results in this 
section and the finding of the previous section. That earlier finding indicated
a low probability that the projected density enhancement on the sky in Aquarius 
would turn out to contain such an elongated structure as the Aquarius 
filament unless there is significant filamentation among Abell clusters. The 
Abell/ACO sample shows no such filamentation. 

\subsection{Dynamics of the Aquarius Knot} 

Knowledge of the masses and dynamical states of the superclusters
discussed in this paper are
significant to the study of the mass distribution on large
scales in general. The entire Aquarius and Aquarius-Cetus filaments each
have spatial extents that imply crossing times that are a few times the 
Hubble time (assuming peculiar velocities of $\sim 1000$ km s$^{-1}$ 
-- see discussion below), so these structures can not have broken away from 
the Hubble Flow and may well not be gravitationally bound overall. Until they have 
been studied in far more detail, little
can be said about dynamics within such extended structures. However, the knot
within the Aquarius filament is dense enough and has sufficient redshift
information available to warrant a closer look now. The same is true of the 
newly-identified Microsopium Supercluster, which we will also examine in this
section. 

The knot within the Aquarius supercluster
is comparable in density enhancement to the Shapley Concentration
and the Corona Borealis (Cor Bor) superclusters (see Section 4.1).
However, Tables 5a and 5b show how sparsely the region has been sampled thus
far. While most of the clusters observed for this program have
enough galaxy redshifts for reliable cluster velocity determinations, none have
enough for accurate velocity dispersions, which are needed for
reasonable cluster mass determinations. 

Recent work by Small et al. (1998) has shown that both the
virial mass estimator and the projected mass estimator
(Bahcall \& Tremaine 1981) give reasonable mass estimates
for clusters with $\sim 30$ or more galaxy observations.
Small et al. (1998) determined a mass for Cor Bor
of at least $3\times 10^{16}h^{-1}M_{\odot}$ and a mass-to-light
ratio of $726h (M/L)_{\odot}$. The small number of galaxies
measured for our initial observations prohibits a similar
determination of mass for the Aquarius superclusters. At best,
a simple sum of the `typical' masses of Abell clusters would
give a lower limit to the total mass. Since both filaments
discovered contain some clusters with single measured redshifts,
only the knot within Aquarius has enough clusters with velocity
dispersions for estimates (although rather rough given the small number
of redshifts available for most of these clusters) of individual
cluster masses. Thus, assuming A2546 and A2555 to be typical Abell/ACO
$R \ge 1$ clusters with $\sigma_v \sim 800$ km s$^{-1}$ (like A2553, A2579, 
and A2554, which according to Zabludoff
 et al. 1990 has $\sigma_v = 827$ km s$^{-1}$ and for which Girardi et al. 1998 
determined a cluster mass of 6.4 $\times 10^{14}h^{-1}M_{\odot}$), we estimate each to have
a cluster mass $\sim 7\times 10^{14}h^{-1}M_{\odot}$ and find a total 
lower-limit
mass for the knot $\sim 6\times 10^{15}h^{-1}M_{\odot}$ (about half
of the mass coming from the large dispersion of A3996). A mean harmonic radius of 
1.5 $h^{-1}$ Mpc was assumed for each virial-theorem estimation, since
that was the average projected distance from the cluster center of the cluster 
members observed. Similar 
calculations using the velocity dispersions reported by Katgert et al. 
(1996) for clusters of the Microscopium supercluster result in a 
lower-limit mass of $\sim 4\times 10^{15}h^{-1}M_{\odot}$ (Girardi et al. 1998
estimate a total mass of 2$\times 10^{14}h^{-1}M_{\odot}$ for three of the 
seven clusters). We
note that the sum of the masses of the clusters in Cor Bor
as determined by Small et al. (1998) in the same manner is $5.3\times 
10^{15}h^{-1}M_{\odot}$,
which is a factor of 6-8 smaller than their mass determinations  
by the virial mass estimator or the projected mass estimator. 

It may soon become feasible to conduct a study of the 
dynamics among these clusters to further constrain
the mass of the system. The 2dF redshift survey of this region, already
underway, should provide a large number of redshifts for galaxies within
and between clusters in the knot (as well as for the rest of the supercluster).
However, another large observational program, employing a secondary distance 
indicator that can provide distances of sufficient accuracy, will be necessary
to determine cluster peculiar velocities in the region. The clusters in the 
knot have a radial distance range of about 13 $h^{-1}$ Mpc (using their 
redshifts directly), at an average distance of about 
325 $h^{-1}$ Mpc, so one would need accuracies of only a few percent
from the secondary indicator. Some standard-candle-type analyses that 
might apply at such distances, such as brightest (or third- or 
tenth-brightest) cluster galaxies, or even entire cluster luminosity 
function analyses as attempted by Small et al. on Corona Borealis, are 
sufficiently coarse that they are unlikely to ever provide the accuracy
needed for a study of dynamics within the knot.
Other methods, such as approaches using relations like Tully-Fisher
or those employing Type I supernovae, hold the promise of achieving 
the necessary accuracy when large amounts of observational data become 
available for their application. 

We can at the present time get a rough conception
of the dynamics of the Aquarius knot and Microscopium by looking at 
typical supercluster
crossing times and virialization timescales and by applying a technique 
Sargent \& Turner (1977) developed to measure
the slowing of the Hubble flow for systems where only
redshift information is known. One can calculate the angle
between the separation vector and the plane of the sky
at the midpoint between clusters using

\begin{equation}
\alpha = \arctan[\frac{1}{2}(z_1/z_2 -1)\cot(\frac{1}{2}\Delta_{12})]; 
~ ~ ~ ~ 0 \le \alpha \le \pi/2,
\end{equation}

\noindent where $z_1 \ge z_2$ and $\Delta_{12}$ is the separation
between clusters 1 and 2 on the sky in radians.
If $\bar{\alpha} < 32.7^{\circ}$ the region is expected to have
a slowed Hubble flow. If violent relaxation has not yet occurred,
such system should appear flattened in the redshift direction. If
$\bar{\alpha} > 32.7^{\circ}$ the region may have reached virial equilibrium
and the system would appear elongated along the line of sight.
If $\bar{\alpha} \sim 32.7^{\circ}$, its isotropic value, we expect
an unperturbed Hubble flow or a system in the midst of relaxation. Using
only a few clusters per supercluster for such calculations clearly will
make for very large error bars, compared to studies of the dynamical states of
clusters or groups of galaxies, but the results below can still provide some
useful indications. 

Postman et al. (1988) calculated
$\bar{\alpha} = 56.5^{\circ} \pm{12.7^{\circ}}$  for Cor Bor
(with A2124 excluded from the analysis because of its apparently greater 
separation from the rest of the supercluster), a significant indication of 
virialization, although they noted that true spatial elongation rather
than dynamical redshift-space elongation could easily be the major factor in 
the case of this one well-studied supercluster.
The value of $\bar{\alpha}$ for the six-cluster
knot within the Aquarius supercluster is 
$\bar{\alpha} = 28.8^{\circ} \pm{18.5^{\circ}}$. Since this result is
well within the errors (determined as in 
Wagner \& Perrenod (1981)) of the isotropic result, the knot is 
clearly not significantly elongated or flattened along the line of
sight. This suggests either that the supercluster has not yet significantly
broken away 
from the Hubble expansion or that it has begun violent relaxation but
is not yet virialized.

From the extent of the system on the sky,
we expect the supercluster crossing time of a typical cluster
with $v_{pec} \sim 800$ km s$^{-1}$ (typical of many recent findings, {\it e. g.}
Bahcall, Gramman \& Cen 1994, Croft \& Efstathiou 1994, but a bit high compared
to $v_{pec} \sim 500$ km s$^{-1}$ from Bahcall \& Oh 1996) to be 
$T_c \sim 5 \times 10^9$ years, thus the Aquarius knot could be gravitationally bound.
The virialization timescale for a spherically
symmetric collapsing mass is (Gunn \& Gott 1972)

\begin{equation}
T_v = \frac{2.14}{\sqrt{G\rho}}
\end{equation}

\noindent where $\rho$ is the current-epoch mass density.
Postman et al. (1988) determined $T_v$ for Cor Bor to
be $\sim 2 \times 10^{10}$ years, so they concluded that it was 
unlikely that the supercluster was virialized, although they found that
the mass of Cor Bor was sufficient to bind the supercluster
(as did Small et al. 1998). 
We would expect $T_v$ to be about the same for the Aquarius knot, which
has roughly the same rich-cluster spatial density as Cor Bor. 
These values of $\bar{\alpha}$,
$T_c$ and $T_v$ for the knot do not allow for discrimination between the 
two cases of continued expansion with the Hubble flow and violent relaxation. 

We also note that $\bar{\alpha}$, $T_c$ and $T_v$ 
are very similar to the values for the Aquarius knot for the seven clusters of 
Microscopium ($\bar{\alpha} = 31.7^{\circ} \pm 17.2^{\circ}$)
and for the nine clusters of Shapley ($\bar{\alpha} = 34.3^{\circ} \pm 
16.2^{\circ}$). Thus, of the four rich-cluster density peaks, none show
noticeable flattening in the redshift direction, indicative of slowing 
expansion. Cor Bor
has a $\bar{\alpha}$ value suggesting virialization (although this seems 
unlikely from other measures), while Aquarius, Microscopium, and Shapley 
have an isotropic appearance, but could possibly be in the process of 
relaxation, given the timescales involved. These results highlight the
importance of further study of all four of these complexes of clusters. Much 
work has already been done on Shapley and Cor Bor, but the Aquarius
and Microscopium peaks could provide new insight into large-scale dynamics
in the universe. 

\subsection{The Other Targeted SCCs} 

The Eridanus SCC was also included in our target strip on the sky (Fig. 1).
The clusters identified in the friends-of-friends analysis as likely members
(A3802, A3817, A3818, A3820, A3834, A3841, and A3845) lie roughly along 
$\theta = 40^{\circ}$ in Fig. 2. However, the clusters near that 
line that we measured (A3820 and A3845 were not observed because they are
slightly fainter than our magnitude limit of $m_{10} = 18.3$ in the strip) 
have widely differing redshifts (see Table 5c), and there 
is no indication of superclustering in that region.
A3817 was observed previously (Batuski et al. 1995).

The redshift distribution that we found for the Grus-Indus SCC (seven clusters 
near $\alpha = 4^h$, also listed 
in Table 5d) is a bit more interesting, with three clusters, A3148, A3166,
and A3171, within the range 
$32800 \le cz \le 37200$ km s$^{-1}$. These clusters have spatial separations 
of 19 and 28 $h^{-1}$ Mpc, suggesting superclustering, but with so few clusters
involved there is no great statistical significance to their proximity. In 100
catalogs of pseudo-clusters distributed randomly in space to a distance 
of 400 $h^{-1}$ Mpc (about 1060 pseudo-clusters per catalog, with exclusion 
of a region bounded by a latitude
limit of $\pm 30^{\circ }$ to simulate galactic obscuration), we found an 
average of 35 `superclusters' per catalog consisting of three 
pseudo-clusters that would percolate at $b = 30 h^{-1}$ Mpc. 
Only five superclusters of six or more points were found per catalog
and a supercluster of 10 or more was only found in one out of five catalogs. 
Thus, with $\sim 10\%$ of clusters in 3-D groupings
of three clusters, an observing program targeting apparent 2-D clumpings on the 
sky would have considerable likelihood of finding some superclusters
of three clusters even if the spatial distribution of the clusters were 
entirely random. On the other hand, finding several more populous 
superclusters, as researchers have done, obviously requires a spatial 
distribution something akin to that represented 
by the two-point spatial correlation functions of Fig. 7.  

\section{Conclusion}

The Aquarius supercluster appears to be a highly significant single structure
of 14 $R \ge 1$ ACO clusters that percolates at $b =  25 h^{-1}$ Mpc
($n = 8\bar{n}$), and extends in a filamentary fashion at least 110 $h^{-1}$ Mpc. 
This is the largest structure involving such rich clusters at such a
density contrast that has been identified to date.
With the inclusion of four 
$R = 0$ ACO clusters that were classed as $R = 1$ by Abell, the apparent 
filament can be traced to an extent of 150 $h^{-1}$ Mpc, with 22 member 
clusters, at $b = 30 h^{-1}$ Mpc.
Aquarius is also the second most
filamentary of the superclusters that percolate among the measured-$z$, rich 
Abell/ACO clusters at $b = 25 h^{-1}$ Mpc, having $R_A = 4.3$ for its ellipsoid fit.
 
The scale of this structure is similar to 
that seen in the Great Wall of 
galaxies (de Lapparent et al. 1988 and 1991) and in the Perseus-Pisces-Pegasus
supercluster ({\it e. g.}, Gregory, Thompson \& Tifft 1981; Haynes et al. 1988; 
and extended with primarily $R = 0$ clusters by Batuski \& Burns 
1985a). The lengths of all these structures are 
approaching 5-10\% of the horizon length of the universe, the scale of many of 
the features observed by COBE in the 
cosmic microwave background (Smoot 1992).

Besides the extent and shape of Aquarius, the high-density peak that it
contains is also of great interest. Our analysis of available redshift data
for the clusters in this peak and the estimated time-scales involved
leaves open the possibility that the grouping may have broken 
away from the Hubble expansion to be currently in collapse toward an eventual
virialized state. Including the seven-cluster supercluster in Microscopium, four 
such large cluster
overdensities are now known to exist within $z\le 0.11$. These
observations constrain the theoretical models for the formation of such 
structure, since simulations based on such models will need to generate 
similar numbers of high density peaks in the cluster distribution, with 
their sizeable impacts on statistics of large-scale structure like the 
two-point spatial correlation function, as illustrated by Postman et al. (1992) 
and Miller et al. (1998). Postman et al. found that Corona 
Borealis accounted for 20\% of the power in $\xi (r)$ for their 156 $R \ge 1$
Abell cluster sample, while Miller et al. found that Cor Bor and
Microscopium contributed $\sim $20\% to $\xi (r)$ in their sample of 289
$R \ge 1$ Abell/ACO clusters.   
These knots of clusters will no doubt also eventually help in the determination
of $\Omega_{\circ }$, once we have secondary distance indicators of sufficient 
accuracy.

The filamentary structure of the Aquarius supercluster may have even greater 
extent than what is reported here. As can be seen in Fig. 3, A2547 and A2550
lie just beyond the high-$z$ end of the filament, in position to extend the 
structure to a length greater than $200 h^{-1}$ Mpc if other clusters (or 
galaxy bridges) are found to fill in the gaps upon future observation.
There are another dozen clusters nearby 
on the sky (mostly fainter than the $m_{10} = 18.9$ limit that was reached
for the small, $10^{\circ } \times 10^{\circ }$ region in the heart of the 
Aquarius clump, but a few with $18.3 < m_{10} < 18.9$, slightly farther
away on the sky) that do not have measured redshifts. These should be observed
soon to see if they might reveal more structure in the region. 

We also report the discovery of the Aquarius-Cetus supercluster,
another prominent filamentary structure among $R \ge 1$ Abell/ACO clusters, 
with its
nine clusters percolating at $n = 8\bar{n}$ and its own apparent knot of four
clusters. (Two of the cluster redshifts for this structure are based on 
single-galaxy measurements, however, and only one of those (A2638) is in a bridging
position that might significantly affect the size and shape of the supercluster
if its redshift is later corrected.) 

The filamentary shapes of both the Aquarius and Aquarius-Cetus superclusters
turn out to be unusual in the spatial distribution of $R \ge 1$ Abell/ACO 
clusters. Our analysis in section 4.2 revealed that, while the Aquarius region
contains two clearly filamentary structures among such rich clusters, the
Abell/ACO samples do not appear to have appreciably more filamentation than
could be expected by chance within a population of objects with similar
two-point correlation function. While poorer clusters and individual galaxies
have been seen to follow filamentary patterns in many observational programs,
the $R \ge 1$ clusters do not appear to participate in such patterns, at least
for aggregates of five or more rich clusters. This result is in disagreement with
the implication from our projected-filament-frequency analysis of Section 4.1 that 
significant filamentation would need to exist in the Abell/ACO sample in order for
a near-line-of-sight filament of the extent of Aquarius to be largely responsible 
the two-dimensional density enhancement seen in that part of the sky.

The superclusters in this extremely interesting region should be observed 
in a very thorough redshift survey 
of thousands of individual galaxies in the immediate
vicinity. (Such a survey is now being conducted as part of the 2dF program
(Jones et al. 1994)).
This is desireable in order to get more velocities of 
cluster members in those cases where only one galaxy has been measured and also 
to look for bridging structure among the clusters, such as that found in 
Chincarini, Rood \& Thompson (1981), de Lapparent et al. (1988 and 1991), 
Gregory \& Thompson (1984), Gregory, Thompson \& Tifft (1981), 
Tarenghi et al. (1979), Postman et al. (1988), Small et al. (1998), and other 
papers on previously-identified superclusters. 
The proximity and filamentary arrangements of so many rich clusters make the 
Aquarius and Aquarius-Cetus superclusters remarkable
occurrences, but given our finding that superclusters of rich Abell/ACO clusters 
show no more than chance elongation, future observations should perhaps be 
expected to show the filaments 
to be a chance alignments of disconnected clumps of clusters, obviating any need
for theoretical models to produce extensive filaments of rich clusters.
The details of possible structure connecting the clusters will 
be important in determining their true significance 
for the purpose of modeling and understanding 
large-scale structure in the universe.




\acknowledgments

DJB, CM, and KAS were 
supported in this work by National Science Foundation Grant AST-9224350.
KAS gratefully acknowledges Sigma Xi, the Scientific Research Society for a 
travel grant for the August 1994 run. CM is thankful for partial support through
a fellowship from Maine Science and Technology Foundation (Grant MSTF 97-25)
and the NASA EPSCOR program.
DJB also wishes to thank the University of Paris 7 and Paris 
Observatory for partial support for this research effort. 
Support was also provided
by CNRS through the Cosmologie GDR Programme. 

The authors performed much of the work for this paper on workstations
obtained under an equipment grant from Sun Microsystems, Inc. to the
University of Maine.

\newpage

\begin{figure}
\plotone{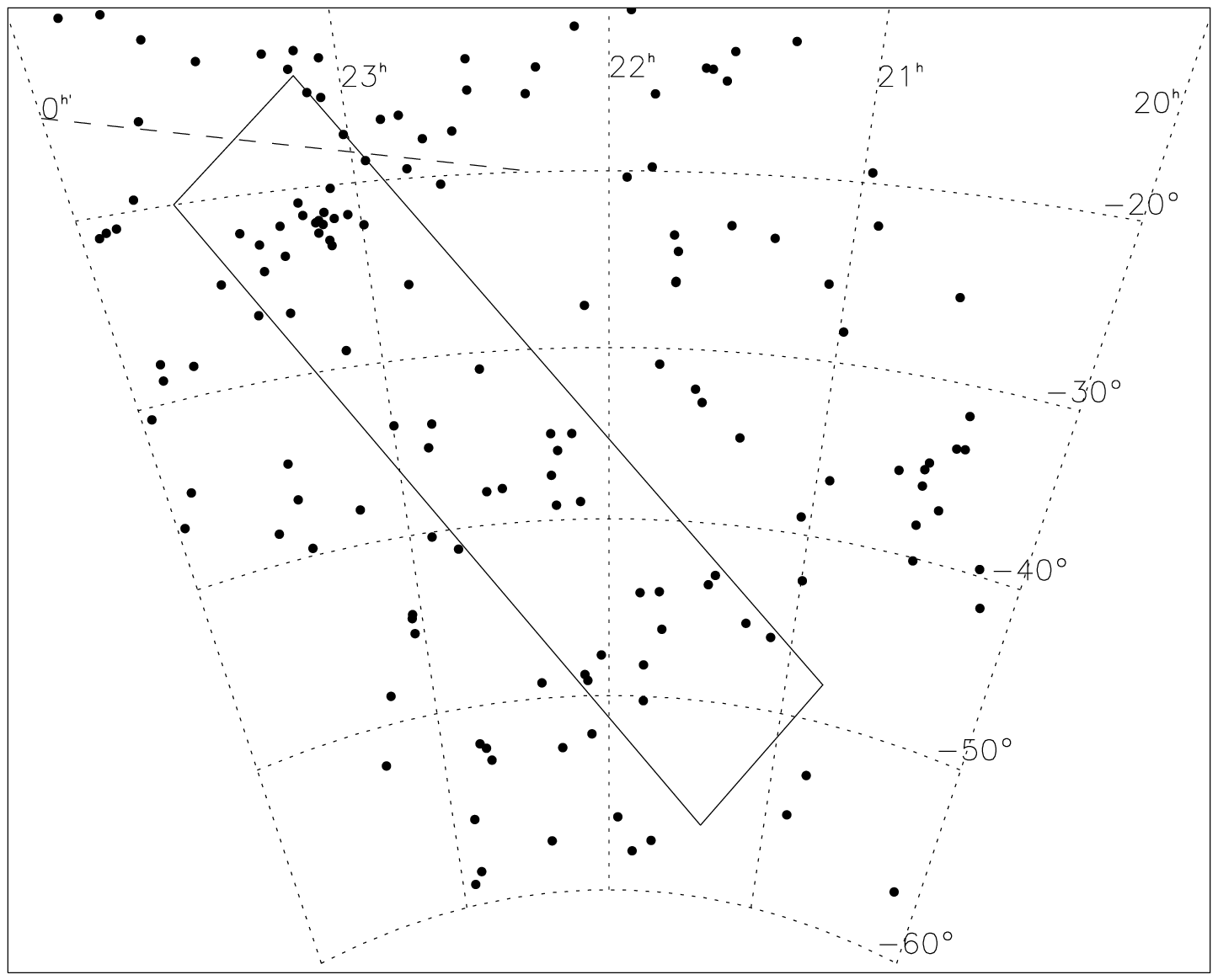}
\caption[]{ A gnomonic projection plot of the $R \ge 1$ ACO
clusters with $m_{10} \le 18.3$ in the general region of the sky of the strip
(outlined by box) containing the two supercluster
candidates targeted for the observations. North of $\delta = -17^{\circ }$
in this plot, the clusters are $R \ge 1$ Abell (1958) clusters and are limited 
to $z \le 0.15$ (measured or estimated), approximately the estimated redshift 
of an $m_{10} \le 18.3$ ACO cluster.}
\end{figure}

\begin{figure}
\epsscale{1.0}
\plottwo{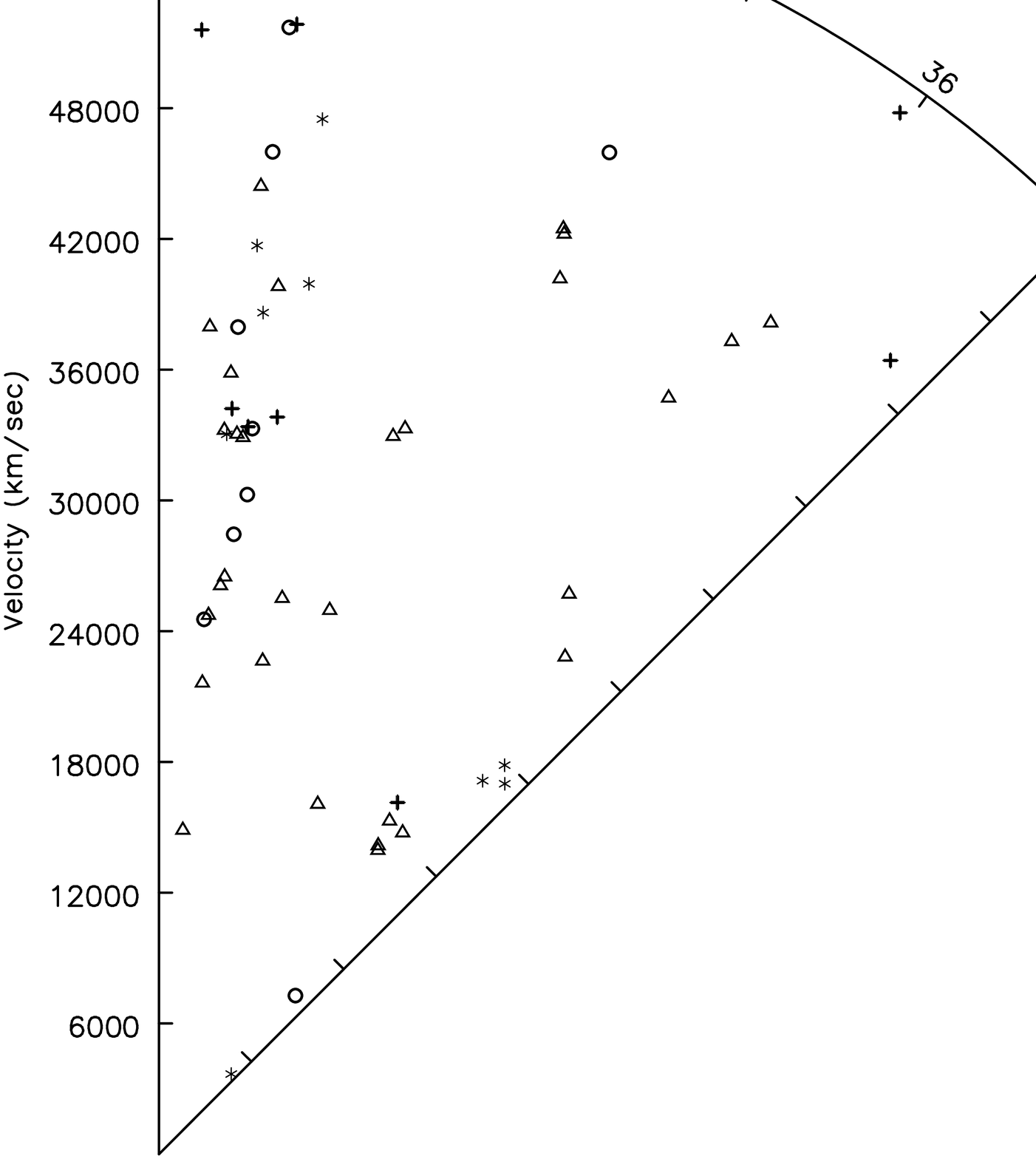}{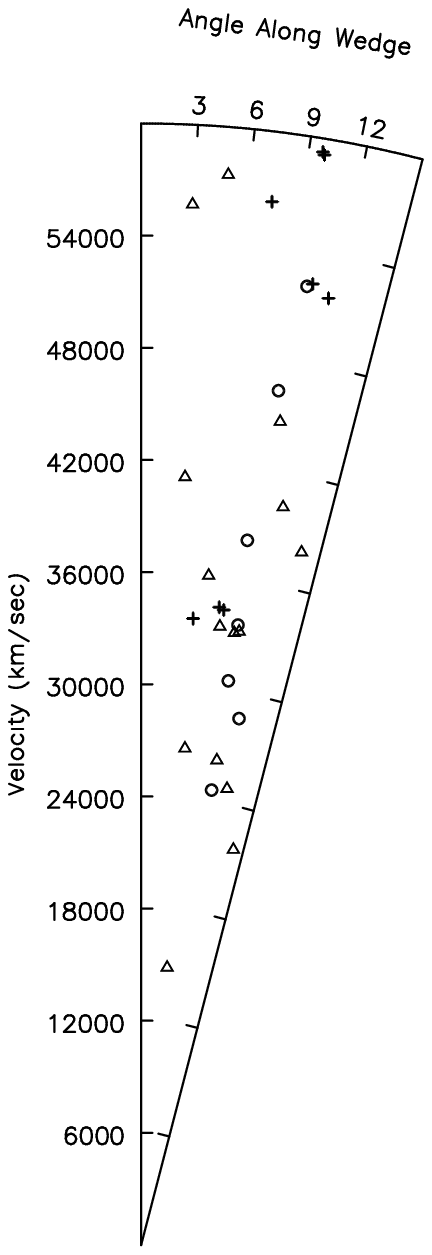}
\caption[]{The left wedge plot shows Abell/ACO $R \ge 0$ cluster positions. The
plot is 45$^{\circ}$ along our observed slice of sky by
10$^{\circ}$ in width (out of
the page). Triangles in this figure represent $R \ge 1$ clusters with 
$m_{10} \le 18.3$ and measured redshifts, which are 87\% complete in the slice.
The crosses are a few fainter $R \ge 1$ clusters, and the asterisks represent
the $R=0$ clusters with measured redshifts in this region.
The right plot is the edge-on view of the left wedge, 
centered on the Aquarius filament and  10$^{\circ}$ in depth
(out of page). This plot includes only $R \ge 1$ clusters, and is 95\%
complete in redshift coverage to $m_{10} = 18.3$. 
The Aquarius supercluster is centered roughly
on $\alpha = 23^h 18^m$ and $\delta = -22^{\circ}$.}
\end{figure}

\begin{figure}
\epsscale{1.0}
\plotone{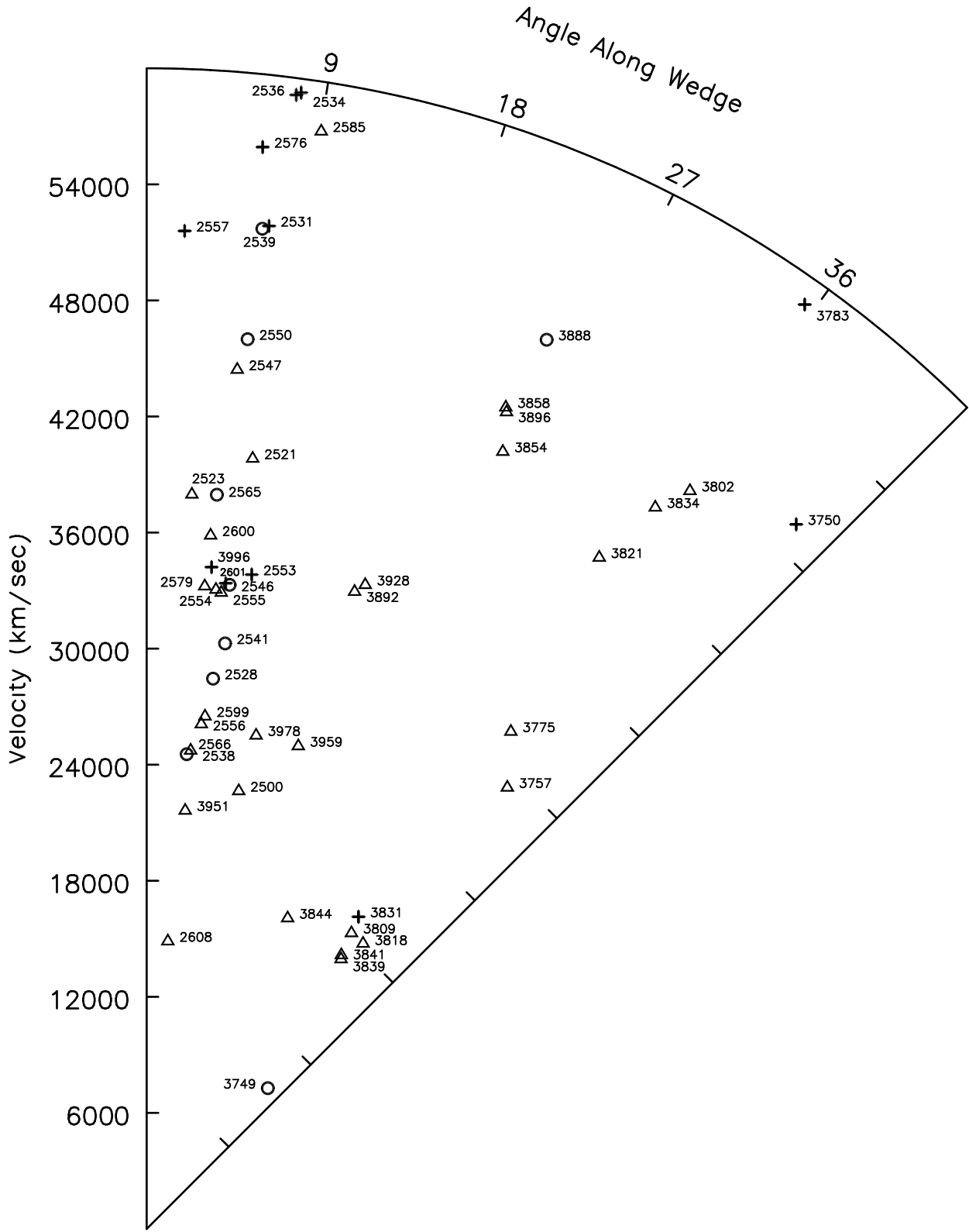}
\caption[]{Similar to Fig. 2 (left), with Abell/ACO cluster
catalog numbers shown. This plot contains only $R \ge 1$ clusters.}
\end{figure}

\begin{figure}
\epsscale{0.5}
\plotone{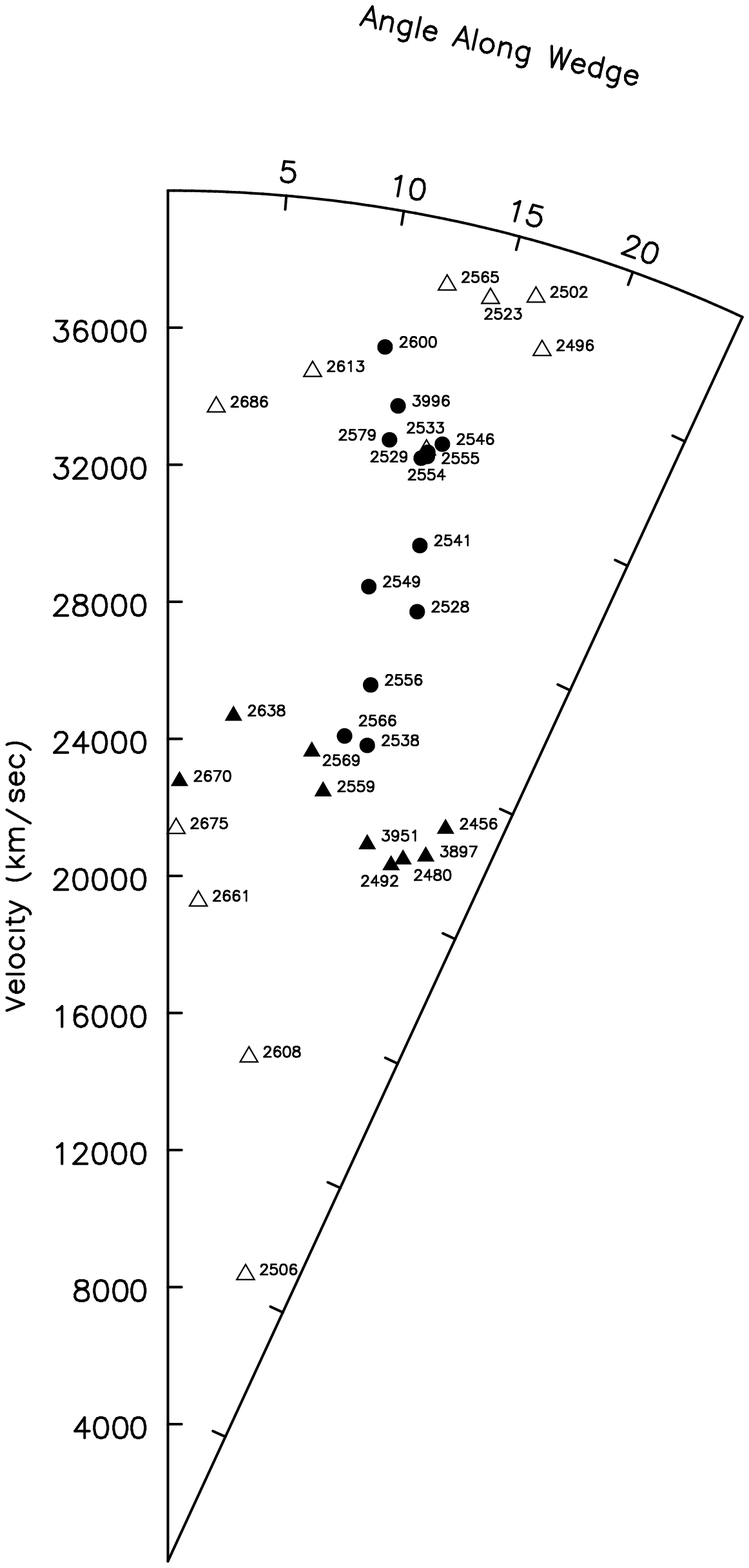}
\caption[]{The centerline for this 25$^{\circ}$ by 12$^{\circ}$ wedge is 
traced by the dashed line in Fig. 1. The filled symbols represent clusters
in the two large superclusters in this region that percolate at 
$n = 8\bar{n}$. Filled triangles represent members of Aquarius-Cetus, and 
filled circles represent those clusters in the Aquarius filament. 
Open triangles stand for clusters that did not percolate into either of 
these two superclusters at $n = 8\bar{n}$.}
\end{figure}

\begin{figure}
\epsscale{0.5}
\plotone{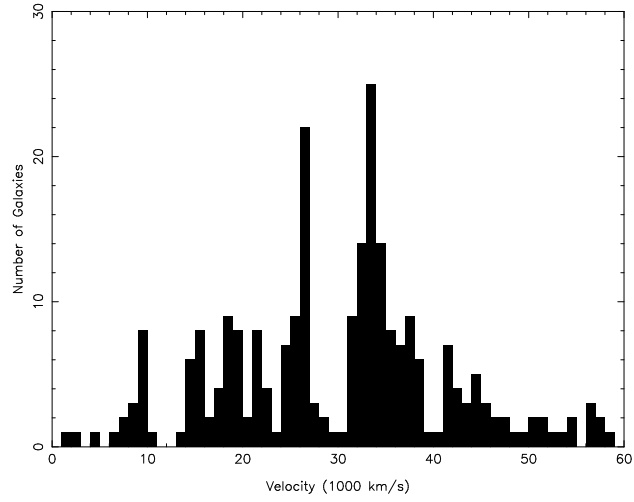}
\caption[]{Histogram of the velocity distribution of 256 galaxies
observed in all 17 cluster fields observed
with MEFOS in the direction of the Aquarius supercluster candidate.
}
\end{figure}

\begin{figure}
\epsscale{0.5}
\plotone{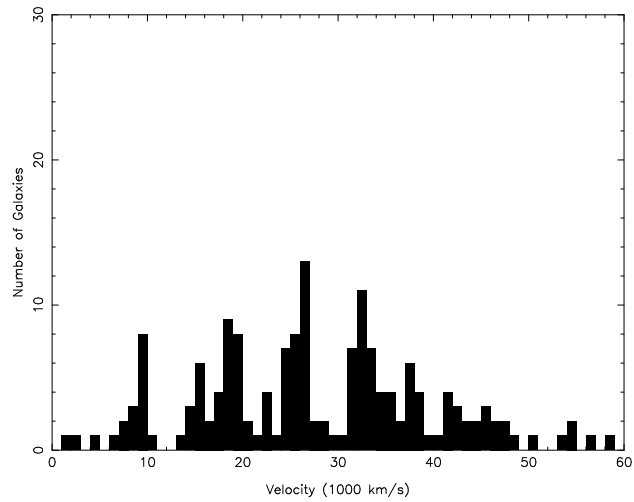}
\caption[]{Histogram of the velocities from the same fields as in Fig. 5,
but with galaxies identified as members of the Abell/ACO clusters
in each observed field subtracted from the sample, leaving 169 galaxy velocities.
}
\end{figure}

\begin{figure}
\plotone{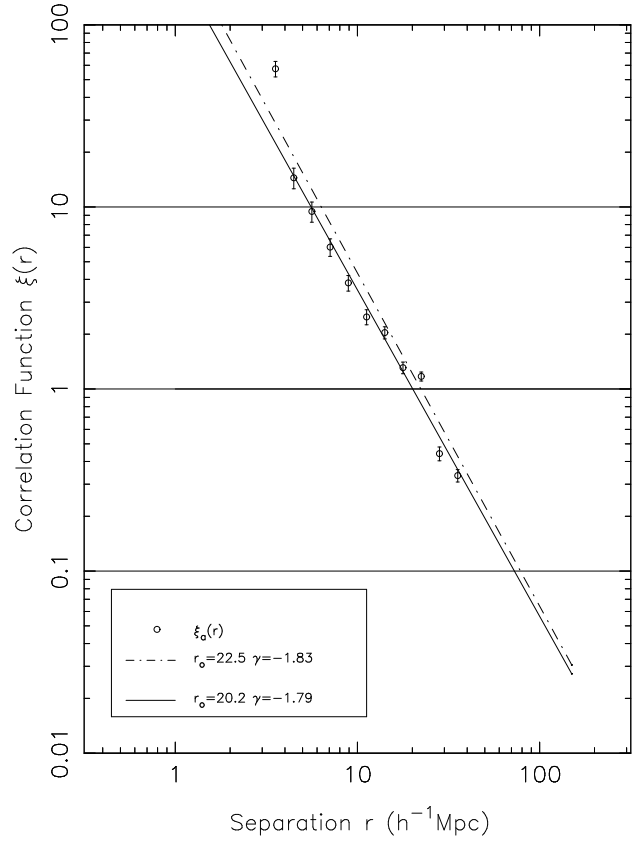}
\caption[]{The two-point correlation function of a single simulated universe 
with cluster positions determined by Equation (3) (solid line), compared with the 
correlation function of Abell clusters (dash-dot line) from Miller et al. (1998).
}
\end{figure}

\end{document}


\tablenum{4}
\begin{deluxetable}{lccccccc}
\tablewidth{0pt}
\tablecaption{Emission Line Galaxy Velocities}
\tablehead{
\colhead{Cluster}         & \colhead{Galaxy}  &
\colhead{Velocity}        & \colhead{Error}  &
\colhead{Lines Found}        \nl 
\colhead{}                & \colhead{Reference}          &
\colhead{(km s$^{-1}$)}       & \colhead{(km s$^{-1}$)} &
\colhead{}}
\startdata
A2557 &  1a & 28074 & 45 & OII\tablenotemark{S/N}\nl 
      &  2a & 19648 & 30 & OII, H${\beta}$\nl 
A2585 &  3a & 26793 & 20 & OII, (H${\beta}$, OIII)\nl 
      &  4a & 13613 & 30 & OII \nl 
      &  5a & 8386  & 30 & OII \nl 
      &  6a & 21308 & 20 & OII, (H${\beta}$)\nl 
      &  7e & 56905 & 20 & OII, H${\beta}$, OIII$_{5007}$, (OIII$_{4958}$) \nl 
A2599 &  8a  & 26839 & 44 & OII \nl 
      &  9a  & 26651 & 20 & OIII{4958}, (H$_{\beta}$), OIII$_{5007}$) \nl 
      & 10e  & 1709  & 20 & H${\beta}$, OIII, (H$_{\gamma}$) \nl 
      & 11e  & 33158 & 20 & OII, H${\beta}$, OIII$_{4958}$, (OIII$_{5007}$) \nl 
      & 12a  & 25938 & 20 & OII, (H${\beta}$) \nl 
      & 13a  & 27213 & 41 & OII\tablenotemark{S/N} \nl 
      & 14a  & 34098 & 20 & OII, H${\beta}$, OIII \nl 
A2600 & 15a  & 18649 & 33 & OII\tablenotemark{S/N} \nl 
      & 16a  & 26094 & 37 & OII \nl 
      & 17a  & 18118 & 20 & OII,  H${\beta}$, OIII \nl 
      & 18a  & 18127 & 20 & OII,  H${\beta}$ \nl 
      & 19a  & 18075 & 40 & OII \nl 
      & 20a  & 18644 & 30 & OII \nl 
      & 21a  & 16389 & 26 & OII, OIII \nl 
      & 22a  & 35358 & 30 & OII \nl 
      & 23a  & 35575 & 30 & OII \nl 
      & 24a  & 34289 & 37 & OII\tablenotemark{S/N} \nl 
      & 25a  & 34522 & 20 & OII, OIII$_{4958}$, (OIII$_{5007}$) \nl 
      & 26e  & 36890 & 30 & OII, (H${\beta}$)\nl 
      & 27a  & 36717 & 31 & OII \nl 
A2601 & 28e  & 37842 & 20 & OII,  H${\beta}$, OIII \nl 
      & 29e  & 2765  & 20 & H${\beta}$, OIII\nl 
      &     &       & & (NeIII, H${\epsilon}$, H$_{\delta}$, H$_{\gamma}$, OIII
${4363}$, HeI) \nl %
      & 30a  & 20201 & 34 & OII, H${\beta}$\nl 
A2608 & 31a  & 35472 & 37 & OII \nl 
      & 32a  & 19579 & 20 & OII, H${\beta}$\nl 
      & 33a  & 14110 & 20 & OII,  H${\beta}$\nl 
A2609 & 34a  & 16504 & 20 & OII, OIII, (H${\beta}$) \nl 
      & 35a  & 27208 & 52 & OII \nl 
      & 36e  & 63277 & 20 & OII,  H${\beta}$, OIII$_{5007}$, (OIII$_{4958}$) \nl 
      & 37a  & 32976 & 20 & OII,  H${\beta}$\nl 
      & 38e  & 25548 & 30 & OII,  H${\beta}$, OIII \nl 
A2641 & 39a  &  9788 & 20 & OII, H${\beta}$, OIII$_{4958}$. (OIII$_{5007}$) \nl 
      & 40e  & 22378 & 30 & OII, OIII${5007}$, (OIII$_{4958}$) \nl 
      & 41e  & 31204 & 20 & OII,  H${\beta}$, OIII \nl 
      & 42a  &  8414 & 20 & H${\beta}$, OIII$_{5007}$, (OII, OIII$_{4958}$) \nl 
      & 43a  & 31675 & 37 & OII \nl 
      & 44a  &  9896 & 20 & OII, OIII${5007}$, (OIII$_{4959}$) \nl 
A3148 & 45a  & 13318 & 23 &  H${\beta}$, OIII \nl 
      & 46e  & 13393 & 36 &  H${\beta}$, OIII \nl 
A3169 & 47a  & 22438 & 34 & OII,  H${\beta}$ \nl 
      & 48e  & 16624 & 30 & OII,  H${\beta}$. OIII \nl 
A3171 & 49e  & 7041  & 35 &  H${\beta}$, OIII \nl 
A3197 & 50a  & 30391 & 26  & OII, OIII \nl 
      & 51a  & 28106 & 24 &  OII,  H${\beta}$, OIII \nl 
A3200 & 52a  & 21276 & 23 & OII,  H${\beta}$, OIII \nl 
A3725 & 53a  & 29832 & 30 & OII,  (H${\beta}$) \nl 
      & 54a  & 12960 & 20 & OII,  H${\beta}$, OIII \nl 
      & 55a  & 12978 & 20 & OII,  H${\beta}$, OIII \nl 
      & 56a  & 16897 & 37 & OII \nl 
      & 57a  & 24764 & 30 & OII \nl 
      & 58a  & 24914 & 20 & OII, H${\beta}$, OIII, (H$_{\gamma}$) \nl 
      & 59a  & 20621 & 31 & OII \nl 
      & 60e  & 32270 & 20 & OII, OIII \nl 
A3750 & 61e  & 49131 & 36 & OII \nl 
      & 62a  & 18288 & 20 & OII,  H${\beta}$, OIII \nl 
      & 63a  & 49610 & 37 & OII \nl 
      & 64e  & 54349 & 44 & OII \nl 
A3757 & 65a  & 2766  & 20 & H$_{\beta}$, OIII, (H$_{\gamma}$, HeI) \nl 
      & 66a  & 23532 & 22 & OII,  H${\beta}$, OIII  \nl 
      & 67a  & 5135  & 20 & H${\beta}$, OIII \nl 
      & 68a  & 4861  & 20 & H${\beta}$, OIII, (NeIII$_{3869}$, HeI) \nl 
      & 69a  & 22678 & 35 & OII, OIII \nl 
      & 70a  & 31429 & 20 & OII,  H${\beta}$, OIII \nl 
A3775 & 74a  & 31944 & 44 & OII \nl 
      & 75a  & 4283  & 20 & NeIII, H${\epsilon}$, H$_{\delta}$, H$_{\gamma}$, H$_{\beta}$, OIII, HeI \nl 
      & 76a  & 19585 & 20 & OII,  H${\beta}$, OIII$_{5007}$, (OIII$_{4958}$) \nl 
      & 77a  &  4930 & 20 &  H${\beta}$, OIII$_{5007}$, (OIII$_{4958}$) \nl 
      & 78a  & 5419  & 20 &  H${\beta}$, OIII$_{5007}$, (OIII$_{4959}$) \nl 
A3834 & 81a  & 10281 & 26 &  H${\beta}$, OIII \nl 
      & 82a  &  8645 & 33 & OIII \nl 
A3839 & 83e  & 17105 & 27 & OII,  H${\beta}$, OIII \nl 
      & 84e  & 60487 & 27 & OII, OIII \nl 
A3841 & 85e  & 17170 & 38 & OII, OIII \nl 
      & 86e  & 15268 & 30 & OII,  H${\beta}$, OIII \nl 
A3844 & 87a  & 17530 & 30 & OII\tablenotemark{S/N} \nl 
      & 88a  & 43925 & 34 & OII\tablenotemark{S/N} \nl 
      & 89a  & 21965 & 30 & OII\tablenotemark{S/N} \nl 
      & 90a  & 26690 & 34 & OII\tablenotemark{S/N} \nl 
      & 91a  & 36303 & 50 & OII\tablenotemark{S/N} \nl 
A3858 & 92e  & 49640 & 31 & OII,  H${\beta}$, OIII \nl 
      & 93e  & 20708 & 32 & OII,  H${\beta}$, OIII \nl 
A3861 & 94e  & 10203 & 24 &  H${\beta}$, OIII \nl 
A3892 & 95a  & 8437  & 24 &  H${\beta}$, OIII \nl 
A3896 & 96e  & 16749 & 38 & OII,  H${\beta}$, OIII \nl 
      & 97a  & 22061 & 24 & OII,  H${\beta}$, OIII \nl 
A3928 & 98a  & 8204  & 23 &  H${\beta}$, OIII \nl 
A3944 & 99a  & 10169 & 33 &  H${\beta}$, OIII \nl 
      & 100a  & 9468  & 22 &  H${\beta}$, OIII \nl 
      & 101a & 30786 & 23 & OII, OIII \nl 
A3951 & 102a & 25295 & 20 & OII, OIII \nl 
      & 103a & 9233  & 29 & OII, H${\beta}$, OIII \nl 
      & 104a & 21524 & 20 & OII, H${\beta}$, OIII \nl 
      & 105a & 54535 & 20 & OII, H${\beta}$, OIII \nl 
      & 106a & 32990 & 26 & OII, H${\beta}$ \nl 
A3959 & 107e & 16419 & 21 & OII, OIII$_{5007}$, (H${\beta}$, OIII$_{4958}$) \nl 
      & 108a & 16480 & 20 & OII,  H${\beta}$, OIII$_{5007}$, (OIII$_{4958}$) \nl 
      & 109e & 26152 & 26 & OII,  H${\beta}$, OIII \nl 
      & 110a & 53016 & 38 & OII,  (H${\beta}$)\nl 
      & 111a & 17735 & 20 & OII,  H${\beta}$, OIII \nl 
      & 112e & 25177 & 65 & OII \nl 
A3978 & 113a & 14864 & 20 & OII,  H${\beta}$, OIII \nl 
      & 114a & 9738  & 20 & OII,  H${\beta}$, OIII$_{5007}$, (OIII$_{4958}$) \nl 
      & 115a & 14952 & 20 & OII,  H${\beta}$, OIII \nl 
      & 116a & 34346 & 22 & OII, OIII$_{4958}$, (OIII$_{5007}$) \nl 
      & 117a & 34924 & 24 & OII \nl 
      & 118a & 14975 & 21 & OII, OIII$_{5007}$, (OIII$_{4958}$) \nl 
A3996 & 119a &  7127 & 26 &  H${\beta}$ \nl 
      & 120e & 32787 & 39 & OII, OIII$_{4958}$, (H${\beta}$, OIII$_{5007}$) \nl 
      & 121a & 17632 & 21 & OII\tablenotemark{S/N},~~~ OIII\tablenotemark{S/N},~~~ ( H${\beta}$)\tablenotemark{S/N} \nl 
      & 122a &  7743 & 20 & OIII, (OII) \nl  
\tablenotetext{S/N}{Low signal to noise}
\tablenotetext{}{Galaxy references with an (a) or an (e) inidcate which
velocity, (a)bsorption or (e)mission, is published in Table 1.}
\tablenotetext{}{Emission lines denoted within parentheses indicate that the
line was detected but 
not included in the velocity determination.}
\enddata
\end{deluxetable}